\documentclass[11pt,a4paper]{article}
\usepackage{amssymb,amsmath}
\numberwithin{equation}{section}
\newtheorem{remark}{Remark}[section]
\newtheorem{proposition}[remark]{Proposition}
\newtheorem{definition}[remark]{Definition}
\newtheorem{theorem}[remark]{Theorem}
\begin{document}
\title{Global solutions for the relativistic Boltzmann equation
in the homogeneous case on the Minkowski space-time}
\author{Norbert Noutchegueme ; Mesmin Erick Tetsadjio T. \\
Department of Mathematics, Faculty of
Science,\\ University of Yaounde I, PO Box 812, Yaounde, Cameroon \\
{e-mail : nnoutch@uycdc.uninet.cm},\\ nnoutch@justice.com}
\date{}
\maketitle
\begin{abstract}
We prove, for the relativistic Boltzmann equation in the
homogeneous case, on the Minkowski space-time, a global in time
existence and uniqueness theorem. The method we develop extends to the cases of
some curved space-times such as the flat Robertson-Walker
space-time and some Bianchi type I space-times.
\end{abstract}
\section{Introduction}

The relativistic Boltzmann equation is one of the basic equations
of the relativistic kinetic theory. An essential tool to describe
the dynamics of a kind of fast moving particles subject to
mutual collisions is their distribution function, denoted by
$f$, and that is a non-negative real-valued function, depending on
the position and the 4-momentum of the particles ; $f$ is
physically interpreted as the "probability of the presence
density" of the particles in a given volume, during their
collisional evolution. In the case of binary and  elastic
collisions, the distribution function $f$ is determined by the
Boltzmann equation, through a non-linear operator called the
"collision operator", that describes, at each point where two
particles collide with each other, the effects of the behavior imposed
by the collision on the distribution function, taking into account
the fact that the momentum of each particle is not the same,
before and after the collision, only the sum of their two momenta
being preserved.

Several authors proved global existence theorems for the non-relativistic
Boltzmann equation and the original result is due to Carleman, in \cite{carleman}.
R. Illner and M. Shinbrot proved that result in \cite{illner}, in the case of small initial data and without assuming symmetry.
An analogous result in the relativistic case is not known. Several authors proved local existence theorems for the
relativistic Boltzmann equation, considering this equation alone,
as Bichteler in \cite{bichteler}, or coupling it to other fields
equations,  as Bancel and Choquet-Bruhat in \cite{bancel1} and
\cite{bancel2}. More interesting would certainly be to look for a
global existence theorem for the relativistic Boltzmann equation. R. T. Glassey and W. Strauss obtained such result in \cite{glassey1}
 for data near to that of an equilibrium solution with non-zero density. It would be interesting to extend to
 the relativistic case the global existence theorem in the case of small initial data.
That was one of the objectives of Mucha, in \cite{mucha1} and
\cite{mucha2}, who studies the relativistic Boltzmann equation
coupled to Einstein's equations, for a flat Robertson-Walker
space-time. Unfortunately, several points in that work are far from clear, namely :

- the formulation of the relativistic Boltzmann equation using
formulae that are valid only in the non-relativistic case ;

- the use of the fixed point theorem without specifying which
complete  metric space is mapped into itself by a contracting map ;

- the use of an approximating operator $Q_{n}$ with unspecified
domain, to approximate the collision operator ;

- the use of an important property of the collision operator, that
requires the symmetry of its kernel, without specifying this
assumption used by Bancel, in \cite{bancel1}.

In this work, we consider the relativistic Boltzmann equation in
the homogeneous case, which means that the distribution function
depends only on the time $t$ and the $4$-momentum $p$ of the
particles. Such cases are very useful for instance in cosmology. In
order to simplify the proofs, we take as background space-time,
the Minkowski space-time, but the techniques we develop easily
extend to the case where the background space-time is the flat
Robertson-Walker space-time or, some Bianchi type I space-time. We
simplify and correct the method followed by Mucha, in \cite{mucha1},
and we prove a global existence theorem for the relativistic
Boltzmann equation whose right hand side is defined by the non-linear
collision operator, whereas the left hand side is defined by a
linear operator. Our method  consists of :

1) constructing an approximation operator $Q_{n}$, with suitable
properties, that will approximate the collision operator, in a suitable
function space ;

2) solving by usual methods, the approximating equation obtained
by replacing the collision operator in the Boltzmann equation by
$Q_{n}$, obtain a global solution $f_{n}$ and prove that $f_{n}$
converges, in a suitable function space, to a global solution of
the relativistic Boltzmann equation.

We begin, first  of all, by giving, following Glassey in
\cite{glassey}, the correct formulation of the relativistic
Boltzmann equation. Next, we construct a suitable framework and an
approximating operator $Q_{n}$, that, in contrast to that
defined in \cite{mucha1}, has a clear domain, and for which we
establish properties, under the symmetry assumption on the kernel
of the collision operator. We then construct a complete metric
space and a contracting map of this space onto itself, that gives,
using the Banach fixed point theorem, the solution $f_{n}$ of the Cauchy
problem for the approximating equation, provided that  the initial
data is sufficiently small, and we show that the solution $f_{n}$
is global. Moreover, we obtain a very simple estimation of the
global solution $f_{n}$ by the initial data that leads to a
considerable simplification of the proofs given in \cite{mucha1}
by showing that there is no need to use, as the author of that paper does,
neither the infinite products, nor the divergence of the numerical
series \ $\sum\frac{1}{n}$, to prove the convergence of $f_{n}$ to a
global solution of the relativistic Boltzmann equation. Finally,
we prove that, if the Cauchy data are sufficiently small, then the
Cauchy problem for the relativistic Boltzmann equation in the
homogeneous case, on the Minkowski space-time, has a unique global
solution $f$ in a suitable function space, that admits a very
simple estimation by the initial data.

The paper is organized as follows : in section 2, we specify the
notations, we define the function spaces, and we introduce the
relativistic Boltzmann equation and the collision operator. We end
this section by a sketch of the strategy adopted to solve the
equation. In section 3, we give the properties of the collision
operator, some preliminary results, and we construct the sequence
of the approximating operators. In Section 4 devoted to the global
existence theorem, we solve the approximating equations, and we
prove the global in time existence theorem.

\section{Notations, function spaces and the relativistic Boltzmann equation}

\subsection{Notations and function spaces}

We denote the Minkowski space-time by $(\mathbb{R}^{4}, \eta)$
with $\eta = Diag(1, -1, -1, -1)$. A Greek index varies from 0 to
3 and a Latin index from 1 to 3. We adopt the Einstein summation
convention :
\begin{displaymath}
A_{\alpha}B^{\alpha}= \sum_{\alpha=0}^{3}A_{\alpha}B^{\alpha}
\end{displaymath}
For $x=(x^{\alpha})=(x^{0}, x^{i})\in \mathbb{R}^{4}$, we set :
\begin{displaymath}
x^{0}=t \ \ ; \ \ \bar{x}=(x^{i}) \ \  ; \ \
\mid \bar{x} \mid = \Big[\sum_{i=1}^{3}(x^{i})^{2}\Big]^{1/2}.
\end{displaymath}
The framework we will refer to is  $L^{1}(\mathbb{R}^{3})$ whose
norm is denoted $\parallel.\parallel$.\\ For $r \in \mathbb{R}$, $ r
> 0$, we set :
\begin{displaymath}
 X_{r} = \{f \in L^{1}(\mathbb{R}^{3}), \ f \geq 0 \
\textrm{a.e.},\ \parallel f \parallel \leq r \}.
\end{displaymath}
Endowed with the distance induced by  $L^{1}(\mathbb{R}^{3})$, $X_{r}$
is a complete and connected metric space.\\
Let $I\subset \mathbb{R}$ be a real interval; we denote by $C[I ;
L^{1}(\mathbb{R}^{3})]$ the Banach space
\begin{displaymath}
C[I ; L^{1}(\mathbb{R}^{3})] = \{f : I \longrightarrow L^{1}(\mathbb{R}^{3}),
f \  \textrm{continuous and bounded} \}
\end{displaymath}
endowed with the norm :
\begin{displaymath}
\mid \parallel f \mid \parallel = \sup_{t \in I}\parallel f(t) \parallel.
\end{displaymath}
We set :
\begin{equation*}
C[I ; X_{r}]= \{f \in C[I, L^{1}(\mathbb{R}^{3})], \ f(t) \in X_r, \ \forall t \in I \}.
\end{equation*}
Endowed with the distance induced by the distance
$d(f, g) = \mid \parallel f-g \mid \parallel$ defined by the norm
$\mid \parallel . \mid \parallel$, $C[I ; X_r]$ is a complete metric space.
We look for the distribution function $f$ of one kind of particles with rest mass $m=1$,
in a collisional evolution, in the space-time $(\mathbb{R}^4, \eta)$ ; $f$ is a non-negative
real-valued function depending on the position $x = (t, \bar{x})$ and the momentum
$p = (p^0, \bar{p}) \in \mathbb{R}^4$ of the particles. $f$ is then defined on the
tangent bundle $T(\mathbb{R}^4)$ that is a 8-dimensional manifold with local coordinates
$(x^\alpha, p^\alpha)$, i.e.
\begin{displaymath}
f : T(\mathbb{R}^4) \longrightarrow \mathbb{R}_+ \ ; \
 (x^\alpha, p^\alpha)\mapsto f(x^\alpha, p^\alpha).
\end{displaymath}
Now the particles are required to move only on the future sheet of the mass shell,
whose equation is $\eta(p, p) = 1$, i.e.
\begin{displaymath}
(p^0)^2 - \sum_{i=1}^3 (p^i)^2 = 1, \ p^0 \geq 0
\end{displaymath}
or equivalently :
\begin{equation}\label{eq:1}
p^0 = \sqrt{1+ |\bar{p}|^2}.
\end{equation}
Hence, $f$ is in fact defined on the 7-dimensional subbundle of $T(\mathbb{R}^4)$
defined by (\ref{eq:1}), and whose local coordinates are $x^\alpha$, $p^i$.
Notice that, in the case of the flat Minkowski space we consider, $\bar{p} = (p^i)$
also stands for the spatial velocity $v= (v^i)$ of the particles.
\subsection{The relativistic Boltzmann equation}

The relativistic Boltzmann equation in $f$, on the flat Minkowski space-time can be written
\begin{equation}\label{eq:2}
p^{\alpha} \frac{\partial f}{\partial x^\alpha} = Q(f,f)
\end{equation}
where $Q$ is the collision operator we now introduce. In the case of binary and
elastic collisions, $Q$ is defined as follows, $p$ and $q$ standing for the momenta
of two particles before their collision, $p'$ and $q'$ for their momenta after the collision,
and where $f$ and $g$ are two functions on $\mathbb{R}^{3}$ :
\begin{verse}
 \begin{equation}\label{eq:3} 1)
 \quad Q(f,g) = Q^+(f,g) - Q^-(f,g)
\end{equation}
with\\
 \begin{equation}\label{eq:4} 2)
 \quad Q^+(f,g) = \int_{\mathbb{R}^{3}}\frac{d\bar{q}}{q^0}\int_{S^2}
  f(\bar{p'})g(\bar{q'})S(\bar{p}, \bar{q}, \bar{p'}, \bar{q'})d\omega
  \end{equation}
 \begin{equation}\label{eq:5} 3)
  \quad Q^-(f,g) = \int_{\mathbb{R}^{3}}\frac{d\bar{q}}{q^0}\int_{S^2}
  f(\bar{p})g(\bar{q})S(\bar{p}, \bar{q}, \bar{p'}, \bar{q'})d\omega
\end{equation}
in which :\\ 4) $S^2$ is the unit sphere of $\mathbb{R}^3$ whose
area element is denoted $d\omega$, and \\5) $S$ is a non-negative
real-valued function of the indicated arguments, called the kernel
of the collision operator $Q$, or the cross section of the
collisions. We suppose that $S$ is bounded i.e.
\begin{equation}\label{eq:6}
  0 \leq S \leq C_1
\end{equation}
where $C_1$ is a positive constant, and we require for $S$ the symmetry assumption :
\begin{equation}\label{eq:7}
S(\bar{p},\bar{q},\bar{p'},\bar{q'}) = S(\bar{p'},\bar{q'},\bar{p},\bar{q}).
\end{equation}
6) As consequences of the conservation law $p+q=p'+q'$, that splits
into
\end{verse}
\begin{equation}\label{eq:8}
  p^{0}+q^{0}=p'^{0}+q'^{0}
\end{equation}
and
\begin{equation}\label{eq:9}
  \bar{p}+\bar{q}=\bar{p'}+\bar{q'}
\end{equation}
 i) we have, using (\ref{eq:1}) and (\ref{eq:8}) :
\begin{equation*}
\sqrt{1+ |\bar{p}|^2} + \sqrt{1+ |\bar{q}|^2} =
\sqrt{1+ |\bar{p'}|^2} + \sqrt{1+ |\bar{q'}|^2}
\end{equation*}
which expresses the conservation of the quantity
\begin{equation}\label{eq:10}
 e = \sqrt{1+ |\bar{p}|^2} + \sqrt{1+ |\bar{q}|^2}
\end{equation}
called the energy of the unit rest mass particles,\\
ii) (\ref{eq:9}) can be expressed, following Glassey in \cite{glassey}, by setting
\begin{equation} \label{eq:11}
\begin{cases}
\bar{p'} = \bar{p} + a(\bar{p}, \bar{q}, \omega)\\
\bar{q'} = \bar{q} - a(\bar{p}, \bar{q}, \omega) \ ; \ \omega \in S^2
\end{cases}
\end{equation}
where $a$ is given by
\begin{equation}\label{eq:12}
  a(\bar{p}, \bar{q}, \omega) = \frac{2e(\Hat{\bar{q}} - \Hat{\bar{p}})
  \sqrt{1+|\bar{p}|^{2}}\sqrt{1+|\bar{q}|^{2}}}{e^{2}-[\omega . (\bar{p}+\bar{q})]^{2}}
\end{equation}
where
\begin{equation*}
 \Hat{\bar{p}} = \frac{\bar{p}}{\sqrt{1+|\bar{p}|^{2}}} = \frac{\bar{p}}{p^{0}}
\end{equation*}
with $e$ defined by (\ref{eq:10}), the dot denoting the usual scalar product in $\mathbb{R}^{3}$,
whereas the jacobian of the change of variables $(\bar{p}, \bar{q})\mapsto (\bar{p'}, \bar{q'})$
defined by (\ref{eq:11}) is
\begin{equation}\label{eq:13}
\frac{\partial(\bar{p'}, \bar{q'})}{\partial(\bar{p}, \bar{q})} =
  - \frac{\sqrt{1+|\bar{p'}|^{2}}\sqrt{1+|\bar{q'}|^{2}}}{\sqrt{1+|\bar{p}|^{2}}\sqrt{1+|\bar{q}|^{2}}}.
\end{equation}

It then appears clearly that, since $\bar{p'}$ and $\bar{q'}$ can be
expressed through (\ref{eq:11}) as functions of $\bar{p}$, $\bar{q}$
and $\omega$, and by (\ref{eq:1}), $q^0$ can be expressed as a function
of $\bar{q}$, the integrals (\ref{eq:4}) and (\ref{eq:5}) that are taken with respect to $\bar{q}$ and
$\omega$, give two functions $Q^{+}(f,g)$, $Q^{-}(f,g)$ of the
single variable $\bar{p}$. In practice, we will consider functions
$f(t)$ defined on $\mathbb{R}^{3}$  with :
\begin{displaymath}
f:I \times \mathbb{R}^{3}\longrightarrow \mathbb{R} \ ; \
(t, \bar{p})\mapsto f(t, \bar{p})
\end{displaymath}
in which we set, for $t$ fixed in the real interval I :
$f(t)(\bar{p})=f(t, \bar{p})$, $ \bar{p}\in \mathbb{R}^{3}$. Also
notice the important fact that, since $\bar{p}$ is not a variable
in the integral (\ref{eq:5}) that defines $Q^{-}$, the operator
$Q^{-}$ has the useful property that :
\begin{equation}\label{eq:14}
Q^{-}(f, g)=fQ^{-}(1, g)
\end{equation}
\begin{remark}
The expression $a(\bar{p}, \bar{q}, \omega) =\omega
.(\bar{p}-\bar{q})$ used by the author in \cite{mucha1} and
\cite{mucha2} is valid only in the non-relativistic case, for very
low velocities ; it should not have been used in the full relativistic
case where fast moving particles, with arbitrary high velocities
that could be not negligible compared to the speed of the light,
are considered.
\end{remark}
We consider the relativistic Boltzmann equation (\ref{eq:2}) in the homogeneous case,
i.e $f$ does not depend on $\bar{x}=(x^{i})$. In this case, (\ref{eq:2}) can be written :
\begin{equation}\label{eq:15}
\frac{df}{dt} = \frac{1}{p^{0}}Q(f,f)
\end{equation}
\begin{remark}
Equations (\ref{eq:2}) and (\ref{eq:15}) do not contain the
derivatives of $f$ with respect to $p^{i}$. This implies that the
dependence of $f$ on $\bar{p}$ will be understood only through the
functional framework chosen. This is where the
difference occurs between the flat case we consider, and the case where
the background space-time could be the flat Robertson-Walker
space-time or, some Bianchi type I space-time, that are curved
space-times. For example, in the flat Robertson-Walker space-time
$(\mathbb{R}^{4},g)$ where the metric $g$ is defined by the line element
:
\begin{displaymath}
ds^{2}=dt^{2}-a^{2}(t)[(dx^{1})^{2}+(dx^{2})^{2}+(dx^{3})^{2}]
\end{displaymath}
where $a > 0$ is a given regular function of $t$, the homogeneous relativistic Boltzmann
equation depends explicitly on the metric and can be written :
\begin{equation}\label{eq:15'}
 \frac{\partial f}{\partial t}-2\frac{\dot{a}}{a} p^{i} \frac{\partial f}{\partial p^{i}} =
  \frac{1}{p^{0}}Q'(f,f)
\end{equation}
where $Q'=a^{3}Q$. Solving  (\ref{eq:15'}) is equivalent to solving
the associated characteristic system that can be written, taking  $t$  as parameter :
\begin{equation}\label{eq:15''}
\frac{dp^{i}}{dt}=-2\frac{\dot{a}}{a}p^{i} \   ;  \   \frac{df}{dt}=\frac{1}{p^{0}}Q'(f,f)
\end{equation}
The equations in $(p^{i})$ being trivial, (\ref{eq:15''}) shows
that the real problem to solve is analogous to (\ref{eq:15}). We
find similar results in the case where the background space-time
is a Bianchi type I space-time whose line element can be written :
\begin{displaymath}
ds^{2}=dt^{2}-a^{2}(t)(dx^{1})^{2}-b^{2}(t)[(dx^{2})^{2}+(dx^{3})^{2}]
\end{displaymath}
with $a>0$, $b>0$ given regular functions of $t$, and that reduces
to the Robertson-Walker line element, when $a=b$ .
\end{remark}
We end this section by indicating the strategy we adopt to solve (\ref{eq:15}).
We consider the following Cauchy problem for (\ref{eq:15}) :
\begin{equation}\label{eq:15'''}
\begin{cases}
\frac{df}{dt} = \frac{1}{p^{0}}Q(f,f)\\
         f(0) =f_{0}
\end{cases}
\end{equation}
where $f_{0}$ stands for the initial data. (\ref{eq:15}) is equivalent the following
integral equation in $f$ :
\begin{equation}\label{eq:16}
f(t,y)=f_{0}(y) +\int_{0}^{t}\frac{1}{p^{0}}Q(f,f)(s,y)ds
\end{equation}
where $y\in \mathbb{R}^{3}$. Our strategy will then consist of :\\
1) constructing an approximating operator $Q_{n}$, with suitable properties that will converge
pointwise in  $L^{1}(\mathbb{R}^{3})$, to the operator $\frac{1}{p^{0}}Q$.\\
2) solving the following approximating integral equation :
\begin{equation}\label{eq:17}
f(t,y)=f_{0}(y) +\int_{0}^{t}Q_{n}(f,f)(s,y)ds
\end{equation}
obtained by replacing in (\ref{eq:16}) $\frac{1}{p^{0}}Q$  by
$Q_{n}$ to obtain a global solution $f_{n}$ that will converge
to a global solution $f$ of (\ref{eq:16}), in a suitable
function space, that will be, given the general framework
adopted, a suitable metric subspace of $C[0 , +\infty ;
L^{1}(\mathbb{R}^{3})]$, namely a space \ $C[0 , +\infty ;X_{r}]$
for a convenient  real number $r>0$.
\section{Preliminary results and approximating operators .}
In this section, we establish some properties of the collision operator $Q$, we construct a
sequence $(Q_{n})$ of approximating operators to $\frac{1}{p^{0}}Q$
and we give some useful properties of $(Q_{n})$.
\begin{proposition}
If $ f , g $ $\in$ $ L^{1}(\mathbb{R}^{3})$,  then
$\frac{1}{p^{0}}Q^{+}(f , g), \frac{1}{p^{0}}Q^{-}(f , g) $ $\in$ $ L^{1}(\mathbb{R}^{3})$ and :\\
\begin{equation}\label{eq:18}
\parallel \frac{1}{p^{0}}Q^{+}(f , g)\parallel \leq C\parallel f\parallel\parallel g\parallel
\end{equation}
\begin{equation}\label{eq:19}
\parallel \frac{1}{p^{0}}Q^{-}(f , g)\parallel \leq C\parallel f\parallel\parallel g\parallel
\end{equation}
\begin{equation}\label{eq:20}
\parallel\frac{1}{p^{0}}Q^{+}(f , f) - \frac{1}{p^{0}}Q^{+}(g , g) \parallel \leq C(\parallel f\parallel + \parallel g\parallel)\parallel f-g \parallel
\end{equation}
\begin{equation}\label{eq:21}
\parallel\frac{1}{p^{0}}Q^{-}(f , f) - \frac{1}{p^{0}}Q^{-}(g , g) \parallel \leq C(\parallel f\parallel + \parallel g\parallel)\parallel f-g \parallel
\end{equation}
\begin{equation}\label{eq:22}
\parallel\frac{1}{p^{0}}Q(f , f) - \frac{1}{p^{0}}Q(g , g) \parallel \leq C(\parallel f \parallel+ \parallel g\parallel)\parallel f-g \parallel
\end{equation}
where $ C=8 \pi C_{1} $, with $ C_{1} $ given by (\ref{eq:6})
\end{proposition}
\underline{Proof}\\
1) The expression (\ref{eq:4}) for  $Q_{+}$  gives, using (\ref{eq:6}) :
\begin{displaymath}
I =\parallel \frac{1}{p^{0}}Q^{+}(f , g)\parallel  \leq C_{1}\int_{\mathbb{R}^{3}} \int_{\mathbb{R}^{3}} \frac{d\bar{p}d\bar{q}}{p^{0}q^{0}} \int_{S^{2}}
\mid f(\bar{p'})\mid  \mid g(\bar{q'})\mid d\omega
\end{displaymath}
Now we consider the change of variables $(\bar{p},\bar{q}) \rightarrow  (\bar{p'},\bar{q'})$
defined by (\ref{eq:11}). The Jacobian (\ref{eq:13}) of that transformation gives,
using (\ref{eq:1}) that :\\
$d\bar{p}d\bar{q} = \frac{p^{0}q^{0}}{(p^{0})'(q^{0})'}d\bar{p'}d\bar{q'}$.
Hence, the above inequality gives, using
$\frac{1}{(p^{0})'(q^{0})'} \leq 1$,
\begin{displaymath}
I\leq C_{1}\int_{\mathbb{R}^{3}}\mid f(\bar{p'})\mid d\bar{p'} \int_{\mathbb{R}^{3}}\mid
 g(\bar{q'})\mid d\bar{q'}\int_{S^{2}}d\omega =4 \pi C_{1}\parallel f \parallel  \parallel g
 \parallel
\end{displaymath}
and (\ref{eq:18}) follows.\\
2) The expression (\ref{eq:5}) for $Q^{-}$ gives, using (\ref{eq:6}) and
$\frac{1}{p^{0}q^{0}} \leq 1$,
\begin{displaymath}
\parallel \frac{1}{p^{0}}Q^{-}(f , g)\parallel
 \leq C_{1}\int_{\mathbb{R}^{3}}\mid f(\bar{p})\mid d\bar{p} \int_{\mathbb{R}^{3}}\mid
 g(\bar{q})\mid d\bar{q}\int_{S^{2}}d\omega =4 \pi C_{1}\parallel f \parallel  \parallel g
 \parallel
\end{displaymath}
and (\ref{eq:19}) follows.\\
3) The inequalities (\ref{eq:20}), (\ref{eq:21}), (\ref{eq:22})
are direct consequences of (\ref{eq:18}), (\ref{eq:19}),
$Q=Q^{+}-Q^{-}$ and the fact that by (\ref{eq:4}), (\ref{eq:5}),
the operators $Q^{+}$, $Q^{-}$ are bilinear and this allows us to
write :
\begin{displaymath}
\frac{1}{p^{0}}Q^{+}(f , f) - \frac{1}{p^{0}}Q^{+}(g , g)=\frac{1}{p^{0}}Q^{+}(f , f-g) + \frac{1}{p^{0}}Q^{+}(f-g , g)
\end{displaymath}
\begin{displaymath}
\frac{1}{p^{0}}Q^{-}(f , f) - \frac{1}{p^{0}}Q^{-}(g , g)=\frac{1}{p^{0}}Q^{-}(f , f-g) + \frac{1}{p^{0}}Q^{+}(f-g , g)
\end{displaymath}
This completes the proof of Proposition 3.1. $\blacksquare$
\begin{remark}
Inequality (\ref{eq:22}) shows that the operator
$\frac{1}{p^{0}}Q$ is locally  lipschitzian in the
$L^{1}(\mathbb{R}^{3})$- norm, with respect to $f$. So, the
standard theorem for ordinary differential equations in
Banach spaces gives the local existence of the solution $f$ of the
Cauchy problem (\ref{eq:15'''}). Moreover, given the conservation
of the $L^{1}$-norm during the evolution, one could deduce the
global existence of the solution $f$ of the Cauchy
problem(\ref{eq:15'''}). But  in our case, the physical background
imposes to take non-negative initial data $f_{0}$ that evolves to
give a non-negative solution $f$ of the Boltzmann equation. This
is why, rather than using the above standard theorem for
ordinary differential equation  that does not give the positivity
property, we develop a method that gives global non-negative
solutions from non-negative initial data, provided that the
initial data is sufficiently small.
\end{remark}
We now state and prove the following result on which relies the
construction of the approximating operators .\\In what follows $C$
is the constant defined in Proposition 3.1.
\begin{proposition}
Let $r_{0}=\frac{1}{7C}$, then , for every $r\in]0 , r_{0}]$, for
every $v\in X_{r}$ and for every integer $n\geq 2$, the equation
\begin{equation}\label{eq:23}
nu -\frac{n}{p^{0}}Q(u , u)= v
\end{equation}
has a unique solution $u_{n}\in X_{r}$.
\end{proposition}
\underline{Proof} : Consider a number $r>0$ and take $v\in X_{r}$ ;
$v$ being fixed . We look for a solution $u$ of (\ref{eq:23}) that
can also be written, using $Q=Q^{+}-Q^{-}$ :
\begin{displaymath}
nu -\frac{n}{p^{0}}Q^{+}(u , u) +  \frac{n}{p^{0}}Q^{-}(u , u)  = v
\end{displaymath}
from which we deduce, using $ Q^{-}(u, u)=uQ^{-}(1, u)$, for $u\geq 0$ a.e.
\begin{displaymath}
   u=\frac{v + \frac{n}{p^{0}}Q^{+}(u , u)}{n + \frac{n}{p^{0}}Q^{-}(1 , u)}
\end{displaymath}
So, $u$ can be considered as a fixed point of the map $F_{n}$ defined by :
\begin{displaymath}
h\longmapsto F_{n}(h)=\frac{v + \frac{n}{p^{0}}Q^{+}(h , h)}{n + \frac{n}{p^{0}}Q^{-}(1 , h)} \qquad(a)
\end{displaymath}
We deduce from (\ref{eq:18}) that $F_{n}:X_{r}\longrightarrow L^{1}(\mathbb{R}^{3})$.
Let us show that one can choose $r>0$ and $n\in \mathbb{N}$ so that $F_{n}$ is a contracting
map from the complete metric space $X_{r}$ into itself. We will then solve the problem by  applying the fixed point theorem.\\
1) We deduce from (a), using (\ref{eq:18}) that, if $h\in X_{r}$, we have since $v\in X_{r}$ :
\begin{displaymath}
\parallel F_{n}(h)\parallel \leq \frac{\parallel v \parallel}{n}+ C \parallel h \parallel\parallel h \parallel  \leq  \frac{r}{n}+ Cr^{2}
\end{displaymath}
which shows that, we have :\\
\begin{center}
$\parallel F_{n}(h)\parallel \leq r$ \ if  \ $n\geq 2$ \ and \ $ 0<r\leq\frac{1}{2C}$ \qquad(b)\\
\end{center}
So (b) gives conditions under which $F_{n}$ maps $X_{r}$ into itself.\\
2) Let $g,h\in X_{r}$ \ and \ $n\geq 2$ \ be given. We deduce from (a) that :\\
\begin{displaymath}
 F_{n}(h)- F_{n}(g)=\frac{v + \frac{n}{p^{0}}Q^{+}(h , h)}{n + \frac{n}{p^{0}}Q^{-}(1 , h)}-\frac{v + \frac{n}{p^{0}}Q^{+}(g , g)}{n + \frac{n}{p^{0}}Q^{-}(1 , g)} ;\qquad(c)
\end{displaymath}
Notice that, since \ $g\geq 0$, $h\geq 0$ a.e. ; we have  :\\
$(n + \frac{n}{p^{0}}Q^{-}(1 , h))(n + \frac{n}{p^{0}}Q^{-}(1 , g)) \geq n^{2}>1$ .
So we deduce from (c) using once more (\ref{eq:14}) i.e the relation : $ Q^{-}(f, g)=fQ^{-}(1, g)$, that : \\
\begin{eqnarray*}
\parallel  F_{n}(h)- F_{n}(g)\parallel \leq \parallel\frac{1}{p^{0}}Q^{+}(h , h) - \frac{1}{p^{0}}Q^{+}(g , g)\parallel
+\parallel\frac{1}{p^{0}}Q^{-}(v , g) - \frac{1}{p^{0}}Q^{-}(v , h)\parallel {} \nonumber\\
+\parallel \frac{1}{p^{0}}Q^{-}[\frac{1}{p^{0}}Q^{+}(h , h) , g] - \frac{1}{p^{0}}Q^{-}[\frac{1}{p^{0}}Q^{+}(g , g),h]\parallel \qquad(d)
\end{eqnarray*}
We can write, using the bilinearity of $Q^{-}$ :
\begin{displaymath}
\frac{1}{p^{0}}Q^{-}(v , g) - \frac{1}{p^{0}}Q^{-}(v , h)=\frac{1}{p^{0}}Q^{-}(v ,g - h).
\end{displaymath}
So, if in (d) we apply (\ref{eq:20}) to the first term and (\ref{eq:19}) to the second term, we obtain
since $\parallel g \parallel \leq r , \parallel h \parallel \leq r$ :
\begin{eqnarray*}
\parallel\frac{1}{p^{0}}Q^{+}(h , h) - \frac{1}{p^{0}}Q^{+}(g , g)\parallel \leq 2Cr\parallel h-g \parallel \\
\parallel\frac{1}{p^{0}}Q^{-}(v , g) -  \frac{1}{p^{0}}Q^{-}(v , h)\parallel \leq Cr\parallel h-g \parallel \qquad (e)
\end{eqnarray*}
Now concerning the third term in (d), we can write, using once
more the bilinearity of $Q^{-}$ :
\begin{eqnarray*}
\frac{1}{p^{0}}Q^{-}[\frac{1}{p^{0}}Q^{+}(h , h) , g] - \frac{1}{p^{0}}Q^{-}[\frac{1}{p^{0}}Q^{+}(g , g),h]=
\frac{1}{p^{0}}Q^{-}[\frac{1}{p^{0}}Q^{+}(h , h) , g-h] + \\
\frac{1}{p^{0}}Q^{-}[\frac{1}{p^{0}}Q^{+}(h , h) -\frac{1}{p^{0}}Q^{+}(g, g), h] \qquad(f)
\end{eqnarray*}
Now we apply :\\
i) to the first term in the right hand side of (f), the property (\ref{eq:19}) of $Q^{-}$
followed by the property (\ref{eq:18}) of $Q^{+}$, \\
ii) to the second term  in the right hand side of (f), the property (\ref{eq:19}) of $Q^{-}$
followed by the property (\ref{eq:20}) of $Q^{+}$.\\ And (f)  gives, since
$\parallel g \parallel \leq r $,\ $\parallel h \parallel \leq r$ :\\
\begin{displaymath}
\parallel\frac{1}{p^{0}}Q^{-}[\frac{1}{p^{0}}Q^{+}(h , h) , g] - \frac{1}{p^{0}}Q^{-}[\frac{1}{p^{0}}Q^{+}(g , g),h]\parallel \leq 3C^{2}r^{2}\parallel g-h \parallel \qquad(g)
\end{displaymath}
Thus (d) gives, using (e) and (g)
\begin{displaymath}
\parallel  F_{n}(h)- F_{n}(g)\parallel \leq 3Cr(1 + Cr)\parallel h-g \parallel  \qquad(h)
\end{displaymath}
So, if we take $r_{0}$=$\frac{1}{7C}$, then for every $r\in]0 , r_{0}]$, we have $3Cr(1 + Cr)< 1/2$
and (b) and (h) show that for every integer $n\geq 2$,
$F_{n}$ is a contracting map from $X_{r}$ into itself. By the fixed point theorem,
$F_{n}$ has a unique fixed point $u_{n}$
that is the unique solution of (\ref{eq:23}).\\ This ends the proof of Proposition 3.3. $\blacksquare$

If $r$ and $n$ are fixed as indicated in Proposition 3.3, to every
$v\in X_{r}$ corresponds a unique $u\in X_{r}$ that satisfies
(\ref{eq:23}). We are then led to the following definition :
\begin{definition}
Let $r_{0}$=$\frac{1}{7C}$. Let $r\in]0 , r_{0}]$ and $n\in \mathbb{N}$, $n\geq 2$ be given.\\
1) Define the operator :
\begin{eqnarray*}
R(n , Q):X_{r} \longrightarrow X_{r} \qquad   u\longmapsto R(n , Q)u
\end{eqnarray*}
as follows : for $u\in X_{r}$, $R(n , Q)u $ is the unique element of $X_{r}$ such that :
\begin{equation}\label{eq:24}
nR(n , Q)u - \frac{n}{p^{0}}Q[R(n , Q)u, R(n , Q)u]= u
\end{equation}
2) Define the operator $Q_{n}$ on $X_{r}$ by
\begin{equation}\label{eq:25}
Q_{n}(u ,u)= n^{2}R(n , Q)u -nu.
\end{equation}
\end{definition}

We give the properties of the operators $R(n , Q)$ and $Q_{n}$ .
\begin{proposition}
Let $r_{0}$=$\frac{1}{7C}$ . Let $r\in]0 , r_{0}]$ and $n, m \in \mathbb{N} , n\geq 2 , m\geq 2$ be given :
Then we have $\forall u ,v\in X_{r} $ :
\begin{equation}\label{eq:26}
\parallel nR(n, Q)u \parallel =\parallel u \parallel
\end{equation}
\begin{equation}\label{eq:27}
\parallel nR(n, Q)u - u \parallel \leq \frac{K}{n}
\end{equation}
\begin{equation}\label{eq:28}
\parallel R(n, Q)u - R(n, Q)v \parallel \leq \frac{2}{n}\parallel u-v \parallel
\end{equation}
\begin{equation}\label{eq:29}
Q_{n}(u, u)=\frac{1}{p^{0}}Q[nR(n, Q)u, nR(n, Q)u]
\end{equation}
\begin{equation}\label{eq:29'}
\parallel Q_{n}(u, u) - Q_{n}(v, v) \parallel \leq K\parallel u-v\parallel
\end{equation}
\begin{equation}\label{eq:30}
\parallel Q_{n}(u, u)-\frac{1}{p^{0}}Q(v, v)\parallel \leq \frac{K}{n}+K\parallel u-v\parallel
\end{equation}
\begin{equation}\label{eq:31}
\parallel Q_{n}(u, u) - Q_{m}(v, v) \parallel \leq K\parallel u-v\parallel+ \frac{K}{n} + \frac{K}{m}
\end{equation}
Here $K=K(r)$ is a continuous function of $r$.
\end{proposition}
\underline{Proof} :
1) (\ref{eq:26}) will be a consequence of the relation :
\begin{equation}\label{eq:32}
\int_{\mathbb{R}^{3}}\frac{1}{p^{0}}Q(f, g)d\bar{p} = 0 \qquad
\forall f,g \in L^{1}(\mathbb{R}^{3})
\end{equation}
we now establish.\\
This is where we need the symmetry assumption (\ref{eq:7})
on the collision \\kernel i.e.
$S(\bar{p}, \bar{q}, \bar{p'},\bar{q'}) = S(\bar{p'},\bar{q'},\bar{p},\bar{q})$.\\
We have, using this relation, the definition (\ref{eq:4}) of \ $Q^{+}$, \ $p^0 = \sqrt{1+ |\bar{p}|^2}$ ,
the change of variables   $(\bar{p},\bar{q}) \rightarrow  (\bar{p'},\bar{q'})$ \ given by (\ref{eq:11}) and
whose Jacobian is given by (\ref{eq:13}) :
\begin{displaymath}
I_{0}=\int_{\mathbb{R}^{3}}\frac{1}{p^{0}}Q^{+}(f , g)d\bar{p}=
\int_{\mathbb{R}^{3}}d\bar{p'}\int_{\mathbb{R}^{3}}d\bar{q'}
\int_{S^2}\frac{ f(\bar{p'})g(\bar{q'})S(\bar{p'}, \bar{q'},
\bar{p}, \bar{p})} {\sqrt{1+ |\bar{p'}|^2 }\sqrt{1+ |\bar{q'}|^2
}}
 d\omega
\end{displaymath}
On the other and, we have, using definition (\ref{eq:5}) of $Q^{-}$ and again\\
$p^0 = \sqrt{1+ |\bar{p}|^2}$ :
\begin{displaymath}
J_{0}=\int_{\mathbb{R}^{3}}\frac{1}{p^{0}}Q^{-}(f , g)d\bar{p}=
\int_{\mathbb{R}^{3}}d\bar{p}\int_{\mathbb{R}^{3}}d\bar{q}
\int_{S^2}\frac{ f(\bar{p})g(\bar{q})S(\bar{p}, \bar{q}, \bar{p'},
\bar{q'})}{\sqrt{1+ |\bar{p}|^2}
 \sqrt{1+ |\bar{q}|^2} }d\omega .
\end{displaymath}
It then appears that \ $I_{0}=J_{0}$ \ and (\ref{eq:32}) follows from
$Q=Q^{+}-Q^{-}$.\\ Given (\ref{eq:32}), (\ref{eq:26}) is obtained by
integrating (\ref{eq:24})
over $\mathbb{R}^{3}$, since $u\geq 0$ a.e. \\
2) The definition (\ref{eq:24}) of $R(n,Q)$ gives, using the
property (\ref{eq:22}) of $Q$ with $f=R(n,Q)u$ and $g=0$:
\begin{displaymath}
\parallel nR(n , Q)u - u \parallel = \parallel \frac{n}{p^{0}}Q[R(n , Q)u, R(n , Q)u]\parallel
\leq Cn(\parallel R(n , Q)u \parallel)^{2}
\end{displaymath}
(\ref{eq:27}) then follows from (\ref{eq:26}) and $\parallel u\parallel\leq r$.\\
3) Subtracting relation (\ref{eq:24}) written for $u$ and $v$ gives :
\begin{displaymath}
R(n, Q)u - R(n, Q)v =\frac{u-v}{n}+\frac{1}{p^{0}}Q[R(n, Q)u, R(n, Q)u]
-\frac{1}{p^{0}}Q[R(n, Q)v, R(n , Q)v]
\end{displaymath}
which gives, using property (\ref{eq:22}) of $Q$ and
(\ref{eq:26}) that gives $\parallel R(n, Q)u \parallel\leq \frac{r}{n}$, \\
$\parallel R(n, Q)v
\parallel \leq \frac{r}{n}$ :
\begin{displaymath}
\parallel R(n , Q)u - R(n , Q)v \parallel \leq \frac{\parallel u-v\parallel}{n}
+\frac{2Cr}{n}\parallel R(n , Q)u - R(n , Q)v \parallel
\end{displaymath}
(\ref{eq:28}) then follows from $n\geq 2$ and the definition of
$r_{0}$ that gives :
$Cr < 1/2$.\\
4) Property (\ref{eq:29}) of $Q_{n}$ is obtained by multiplying  (\ref{eq:24}) by $n$,
using the bilinearity of the operator $Q$, the definition (\ref{eq:25})  of $Q_{n}$.\\
5) Property (\ref{eq:29'}) is obtained by using (\ref{eq:29}),
property (\ref{eq:22}) of $Q$, followed
by properties (\ref{eq:26}) and (\ref{eq:28}) of $R(n,Q)$.\\
6) Notice that, by (\ref{eq:26}), $nR(n,Q)u\in X_{r}$ if $u\in X_{r}$. Then use the property
(\ref{eq:29}) of $Q_{n}$, the property (\ref{eq:22}) of $Q$ to obtain :
\begin{displaymath}
\parallel Q_{n}(u ,u)-\frac{1}{p^{0}}Q(v , v)\parallel \leq 2Cr
\parallel nR(n , Q)u - v \parallel \leq 2Cr
(\parallel nR(n , Q)u - u \parallel +\parallel u-v \parallel)
\end{displaymath}
Hence, (\ref{eq:27}) gives (\ref{eq:30}).\\
7) (\ref{eq:29}) gives :
\begin{displaymath}
Q_{n}(u ,u) - Q_{m}(v ,v) = \frac{1}{p^{0}}Q[nR(n,Q)u, nR(n,Q)u)]
-\frac{1}{p^{0}}Q[mR(m,Q)v, mR(m,Q)v].
\end{displaymath}
then, (\ref{eq:31}) follows(\ref{eq:22}), (\ref{eq:26})
and, (\ref{eq:27}), adding and subtracting $u$ and $v$ .\\ This ends
the proof of Proposition 3.5. $\blacksquare$
\begin{remark}
1) (\ref{eq:30}) shows, by taking $u=v$, that \ $Q_{n}$ converges pointwise
in the $L^{1}(\mathbb{R}^{3})$-norm to the operator $\frac{1}{p^{0}}Q$. From there,
the terminology of "approximating operators " we give to the sequence $Q_{n}$.\\
2) In  \cite{mucha1}, the author defined the operator $R(n,Q)$
whose domain is $X_{r}$ using the equation
$nu-\frac{1}{{p}^{0}}Q(u,u)= v$, instead of (\ref{eq:23}) and he
associated to it, the approximating operator
$Q_{n}(u,u)=nR(n,Q)nu-nu$. But it is clear that, even if $u\in
X_{r}$, for $n$ sufficiently large, $nu$ is no longer in $X_{r}$;
so the domain of this operator \ $Q_{n}$ is unspecified,
in contrast to \ $Q_{n}$ defined by (\ref{eq:25}), for which such a problem does not exist.\\
3) The useful relation (\ref{eq:32}) was also used in \cite{mucha1}, but without specifying
the symmetry assumption (\ref{eq:7}) on the collision kernel $S$, on which this result is based.
\end{remark}
We now have all the tools we need to prove the global existence theorem for the homogeneous
relativistic Boltzmann equation.
\section{The global existence theorem}
With the operator $Q_{n}$ defined above, we will first state and
prove a global existence theorem for the equation (\ref{eq:17}) we
call "approximating equation". Next we establish that the solution
$f_{n}$ of (\ref{eq:17}) converges, in a sense to be specified, to a
global solution of (\ref{eq:16}). We prove these results under the
smallness assumption on the initial data .
\begin{proposition} \label{p:4.1}
Let $r_{0}=\frac{1}{7C}$. Let $r\in ]0, r_{0}]$, $g_{0}\in X_{r}$, $n\in \mathbb{N}$, $n\geq 2$
be given. Then, for every $t_{0}\in [ 0, +\infty[$, the integral equation :
\begin{equation}\label{eq:33}
f(t_{0}+t, y)= g_{0}(y)    +\int_{t_{0}}^{t_{0}+t}Q_{n}(f,f)(s,y)ds,\qquad t\geq 0
\end{equation}
has a unique global solution $f_{n}\in C[ t_{0},+\infty ; X_{r}]$.
Moreover, $f_{n}$ satisfies the inequality :
\begin{equation}\label{eq:34}
\mid \parallel f_{n} \mid \parallel \leq \parallel g_{0}\parallel
\end{equation}
\end{proposition}
\underline{Proof} :
We give the proof in two steps.\\
\underline{Step 1} : Local existence and Estimation.

a) Local existence \\
Let us consider equation (\ref{eq:33}) on an interval $[t_{0}, t_{0}+ \delta]$, $\delta > 0$, i.e.
\begin{equation}\label{eq:35}
f(t_{0}+t, y)= g_{0}(y)    +\int_{t_{0}}^{t_{0}+t}Q_{n}(f,f)(s,y)ds,\quad
t\in [0,\delta],\quad \delta >0
\end{equation}
The relations :
\begin{eqnarray*}
\frac{d}{dt}[e^{nt}f(t_{0}+t, y)]=e^{nt}[\frac{df}{dt}+nf](t_{0}+t,y) ;\\
Q_{n}(f, f)+ nf =n^{2}R(n, Q)f \qquad [Cf\ \ (\ref{eq:25}) ]
\end{eqnarray*}
show that (\ref{eq:35}) is equivalent to :
\begin{equation}\label{eq:36}
f(t_{0}+t, y)=e^{-nt}g_{0}(y) +\int_{0}^{t}n^{2}e^{-n(t-s)}R(n, Q)f(t_{0}+s,y)ds
\end{equation}
we solve (\ref{eq:36}) by the fixed point theorem .\\
Consider the operator $A$ defined on $C[t_{0},t_{0}+\delta ; X_{r}]$ by the
right hand side of (\ref{eq:36}), i.e
\begin{equation}\label{eq:37}
Af(t_{0}+t,y)=e^{-nt}g_{0}(y) +\int_{0}^{t}n^{2}e^{-n(t-s)}R(n, Q)f(t_{0}+s,y)ds
\end{equation}
Let us show that one can find $\delta > 0$ such that $A$ is a contracting map of
the complete metric space $C[t_{0},t_{0}+\delta ; X_{r}]$ into itself .\\
i)Let $f \in C[t_{0},t_{0}+\delta ; X_{r}]$ . Since $ \parallel g_{0}\parallel \leq r$
and using (\ref{eq:26}) that gives \\
$\parallel nR(n, Q)f(t_{0}+s)\parallel = \parallel f(t_{0}+s)
\parallel \leq r$, (\ref{eq:37}) gives for every $t \in
[0,\delta]$ :
\begin{displaymath}
\parallel Af(t_{0}+t)\parallel \leq e^{-nt}r +nre^{-nt}\int_{0}^{t}e^{ns}ds
\leq e^{-nt}r + re^{-nt}[e^{nt}-1] = r
\end{displaymath}
Then $\mid\parallel Af \mid\parallel \leq r$ and this shows that
$A$ maps
$C[t_{0},t_{0}+\delta ;X_{r}]$ into itself .\\
ii) Let $f, g \in C[t_{0},t_{0}+\delta ; X_{r}]$. (\ref{eq:37}) gives :
\begin{displaymath}
(Af-Ag)(t_{0}+t, y)=\int_{0}^{t}n^{2}e^{-n(t-s)}[R(n, Q)f - R(n, Q)g](t_{0}+s,y)ds
\end{displaymath}
which gives, using property (\ref{eq:28}) of $R(n, Q)$ and $e^{-n(t-s)}\leq 1$ :\\
\begin{displaymath}
\mid\parallel Af-Ag \mid\parallel \leq 2n\delta \mid\parallel f-g \mid\parallel
\end{displaymath}
It then appears that $A$ is a contracting map in any space $ C[t_{0},t_{0}+\delta ;X_{r}]$
where $2n\delta\leq 1/2$ i.e $\delta \in ]0, \frac{1}{4n}]$. Taking \ $\delta = \frac{1}{4n}$,
we conclude that $A$ has a unique fixed point
$f_{n} ^{0}\in C[t_{0},t_{0}+ \frac{1}{4n} ; X_{r}]$,
that is the unique solution of (\ref{eq:36}) and hence, the unique solution of (\ref{eq:35}).

b) Estimation\\
Notice that (\ref{eq:36}) gives, taking $t=0, f(t_{0},y)=g_{0}(y)$ ; $f_{n}^{0}$ \ then satisfies :
\begin{displaymath}
f_{n}^{0}(t_{0}+t, y)=e^{-nt}f_{n}^{0}(t_{0}, y) +\int_{0}^{t}n^{2}e^{-n(t-s)}R(n, Q)f_{n}^{0}(t_{0}+s,y)ds
\end{displaymath}
which gives, multiplying by $e^{nt}$, using once more (\ref{eq:26}), and for every \\
$t\in ]0, \frac{1}{4n}]$ :
\begin{displaymath}
 \parallel e^{nt}f_{n}^{0}(t_{0}+t)\parallel \leq \parallel f_{n}^{0}(t_{0})\parallel+
 n\int_{0}^{t}\parallel e^{ns} f_{n}^{0}(t_{0}+s)\parallel ds
\end{displaymath}
this gives by Gronwall's Lemma :
$e^{nt}\parallel f_{n}^{0}(t_{0}+t)\parallel \leq e^{nt} \parallel f_{n}^{0}(t_{0})\parallel$ ;
hence ;
\begin{equation}\label{eq:38}
\mid\parallel f_{n}^{0}\mid\parallel \leq \parallel f_{n}^{0}(t_{0})\parallel
\end{equation}
\underline{Step 2} : Global existence, Estimation and Uniqueness .\\
Let $k \in \mathbb{N}$. Taking in (\ref{eq:35})
$\delta=\frac{1}{4n}$ and replacing in that integral equation,
$t_{0}$ by : $t_{0}+\frac{1}{4n}$; $t_{0}+\frac{2}{4n}$; $\cdots$;
$t_{0}+\frac{k}{4n}$; $\cdots$, step 1 tells us that on each
interval $I_{k}=[t_{0}+\frac{k}{4n},
t_{0}+\frac{k}{4n}+\frac{1}{4n}]$ whose length is $\frac{1}{4n}$,
the initial value problem for the corresponding integral equation
has a unique solution $f_{n}^{k}\in C[I_{k};X_{r}]$ provided that
the initial data we denote $f_{n}^{k}(t_{0}+\frac{k}{4n})$, is a
given element of $X_{r}$; $f_{n}^{k}$ \ then satisfies :
\begin{eqnarray*}
\begin{cases}
f_{n}^{k}(t_{0}+\frac{k}{4n}+t,y)=
f_{n}^{k}(t_{0}+\frac{k}{4n},y)+
\displaystyle\int_{t_{0}+\frac{k}{4n}}^{t_{0}+\frac{k}{4n}+t}Q_{n}(f_{n}^{k},f_{n}^{k})(s, y) ds\ \\
f_{n}^{k}(t_{0}+\frac{k}{4n}) \in X_{r}, \ k\in\mathbb{N}, \ t\in [0, \frac{1}{4n}]
\end{cases}
\end{eqnarray*}
and (\ref{eq:38}) implies :
$\mid\parallel f_{n}^{k} \mid\parallel\leq  \parallel f_{n}^{k}(t_{0}+\frac{k}{4n}) \parallel$ \\
Notice that :
\begin{displaymath}
[t_{0}+ \infty[ = \bigcup_{k \in \mathbb{N}} \left[t_{0}+\frac{k}{4n},t_{0}+\frac{k+1}{4n}\right]
\end{displaymath}
and define :
$f_{n}^{k}(t_{0}+\frac{k}{4n},y)=f_{n}^{k-1}(t_{0}+\frac{k}{4n},y), \ \textrm{if} \
\  k\geq1$ \ and \ $f_{n}^{0}(t_{0})=g_{0}$.\\ Then, the solutions \ $f_{n}^{k-1}$, $f_{n}^{k}$ \ that
are defined on \ $[t_{0}+\frac{k-1}{4n},t_{0}+\frac{k}{4n}] $\\ and \ $[t_{0}+\frac{k}{4n},t_{0}+\frac{k+1}{4n}] $
overlap at $t=t_{0}+\frac{k}{4n}$.\\
Define the function $f_{n}$ on $[t_{0}, + \infty[$ by :
$f_{n}(t)=f_{n}^{k}(t)$  if \ $t \in [t_{0}+\frac{k}{4n},t_{0}+\frac{k+1}{4n}]$\\
then, a straightforward calculation shows, using the above relations for\\
$k, k-1, k-2, \cdots, 1, 0$, that, $f_{n}$ is a global solution of
(\ref{eq:33}) on $[t_{0}, + \infty[$, that $f_{n}$ satisfies the
estimation (\ref{eq:34}) and hence
$f_{n}\in C[t_{0}, + \infty; X_{r}[$ .\\
Now suppose that $f,g \in C[t_{0}, + \infty ; X_{r}]$ are two solutions of (\ref{eq:33}), then :\\
\begin{displaymath}
f(t_{0}+t, y)-g(t_{0}+t, y)=\int_{0}^{t}[Q_{n}(f, f)-Q_{n}(g, g)](t_{0}+s, y)ds ,\quad t\geq0
\end{displaymath}
this implies, using the property (\ref{eq:29'}) of $Q_{n}$ that :\\
\begin{displaymath}
\parallel (f-g)(t_{0}+t)\parallel \leq K\int_{0}^{t}\parallel (f-g)(t_{0}+s)\parallel ds \quad t\geq0
\end{displaymath}
which implies, by the Gronwall's Lemma, that $f=g$ and the uniqueness is proved.
This ends the proof of Proposition \ref{p:4.1}.  $\blacksquare$
\begin{remark}
Equation (\ref{eq:17}) is the particular case of equation (\ref{eq:33})
when $t_{0}=0$, $g_{0}=f_{0}$.
\end{remark}
We now state and prove the global existence theorem.
\begin{theorem}\label{t:4.3}
Let $ r_{0}=\frac{1}{7C}$. Let $ r\in ]0,r_{0}]$ and $ f_{0}\in X_{r}$ be given.
Then the Cauchy problem for the homogeneous Boltzmann equation on the Minkowski
space, with initial data $f_{0}$, has a unique global solution
$f\in C[0, +\infty ; X_{r}] $ ; f satisfies the estimation :
\begin{equation}\label{eq:39}
\begin{displaystyle}
\sup_{t\in [0, +\infty [}\parallel f(t)\parallel \leq \parallel f_{0}\parallel.
\end{displaystyle}
\end{equation}
\end{theorem}
\underline{Proof} : We give the proof in two steps.\\
\underline{Step 1} : Local existence and Estimation\\
Let $t_{0}\in \mathbb{R}$, $g_{0}\in X_{r}$ \ be given, the proposition \ref{p:4.1}
gives for every \ $n \in \mathbb{N}$, $n \geq 2,$ \ the existence of a solution
$f_{n}\in C[t_{0}, +\infty ; X_{r}]$ of (\ref{eq:33}) that
satisfies (\ref{eq:34}). It is important to notice that this
solution depends on $n$.\\
Let $T>0$ be given ; we have $f_{n}\in C[t_{0},t_{0}+T ; X_{r}] \ \forall n\in\mathbb{N}$,
$n\geq 2$. Let us prove that the sequence $(f_{n})$ converges in $C[t_{0},
t_{0}+T ; X_{r}]$  to a solution of the integral equation :
\begin{equation}\label{eq:40}
f(t_{0}+t, y)=
g_{0}(y)+\int_{t_{0}}^{t_{0}+t}\frac{1}{p^{0}}Q(f,f)(s,y)ds,\qquad
t\in [0, T]
\end{equation}
Consider two integers $n, m\geq 2$. We deduce from (\ref{eq:33}) that $\forall t\in [0, T]$ :
\begin{displaymath}
f_{n}(t_{0}+t, y)-f_{m}(t_{0}+t, y)=\int_{t_{0}}^{t_{0}+t}[Q_{n}(f_{n}, f_{n})-Q_{m}(f_{m}, f_{m})](s,y)ds
\end{displaymath}
this gives using the property (\ref{eq:31}) of the operators \ $Q_{n}$, $Q_{m}$ :
\begin{displaymath}
\parallel f_{n}(t_{0}+t)-f_{m}(t_{0}+t)\parallel \leq(\frac{K}{n}+\frac{K}{m})T
+K\int_{0}^{t}\parallel (f_{n}-f_{m})(t_{0}+s)\parallel ds
\end{displaymath}
which gives, using Gronwall's Lemma :
\begin{displaymath}
\mid\parallel f_{n}-f_{m}\mid\parallel
\leq(\frac{1}{n}+\frac{1}{m})KT e^{KT}
\end{displaymath}
this prove that $f_{n}$
is a Cauchy sequence in the complete metric space \\ $C[t_{0},t_{0}+T ;X_{r}]$. Then,
there exists $f\in C[t_{0},t_{0}+T ; X_{r}]$ such that
\begin{equation}\label{eq:41}
f_{n} \ \textrm {converges to} \ f \ \textrm {in} \ C[t_{0}, t_{0}+T ; X_{r}]
\end{equation}
Let us show that $f$ satisfies (\ref{eq:40}). (\ref{eq:41}) implies that :
$f_{n}(t_{0}+t)$ converges to $f(t_{0}+t)$ in $X_{r}$
for every $t\in [0, T]$ ; this implies in particular, by taking $t=0$, and
since $f_{n}(t_{0})=g_{0}$ that $f(t_{0})=g_{0}$. Next we have, using property (\ref{eq:30})
of the operators \ $Q_{n}$ and $Q$ :
\begin{displaymath}
\parallel\int_{t_{0}}^{t_{0}+t}[Q_{n}(f_{n},f_{n})-\frac{1}{p_{0}}Q(f, f)](s, y)ds\parallel
\leq\frac{KT}{n} +KT\mid\parallel f_{n}-f \mid\parallel
\end{displaymath}
which shows that :
\begin{displaymath}
\int_{t_{0}}^{t_{0}+t}Q_{n}(f_{n},f_{n})(s)ds  \ \
\textrm{converges to} \int_{t_{0}}^{t_{0}+t}\frac{1}{p_{0}}Q(f,
f)(s)ds  \ \ \textrm{in } L^{1}(\mathbb{R}^{3})
\end{displaymath}
hence, since $f_{n} $ satisfies (\ref{eq:33}), $f$ satisfies
(\ref{eq:40}) $\forall t\in [0,T]$
and a.e.with respect to $y\in \mathbb{R}^{3}$.\\ Finally (\ref{eq:34}) and (\ref{eq:41})
imply
\begin{equation}\label{eq:42}
\mid\parallel f \mid\parallel \leq \parallel f(t_{0})\parallel
\end{equation}
\underline{Step 2} : Global existence, Estimation and Uniqueness .\\
Since $t_{0}\in [0,+\infty [$ is arbitrary and since $\forall
n\geq 2$, the solution $f_{n}$ of (\ref{eq:33}) is globally
defined on $[t_{0},+\infty[$, Step 1 tells us that, given $T>0$, by
taking in the integral equation (\ref{eq:40}) : $t_{0}=0$,
$t_{0}=T$, $t_{0}=2T$, $\cdots$, $t_{0}=(k-1)T$, $t_{0}=kT$, $\cdots$,
$k\in\mathbb{N}$, then, on each interval $J_{k}=[kT,(k+1)T]$ whose
length is $T$, the initial value problem for the corresponding
integral equation has a solution $f^{k}\in
C[J_{k};X_{r}]$, provided that the initial data we denote
$f^{k}(kT)$ is a given element of $X_{r}$. We then proceed
exactly as in Step 2 of the proof of Proposition \ref{p:4.1} by
writing for $T>0$ \ given, that :
\begin{displaymath}
[0, + \infty[ = \bigcup_{k \in \mathbb{N}} \ [kT, (k+1)T],
\end{displaymath}
to overlap the local solutions $f^{k}\in C[kT, (k+1)T; X_{r}]$, $k
\in\mathbb{N}$ \ with \\ $f^{0}(0)=f_{0}$ and obtain a global solution
$f\in C[0,+ \infty;X_{r}]$ of the equation :
\begin{equation}\label{eq:43}
f(t, y)=f_{0}(y)+\int_{0}^{t}\frac{1}{p_{0}}Q(f, f)(s, y)ds
\end{equation}
that satisfies, using (\ref{eq:42}) that gives $\mid\parallel
f^{k} \mid\parallel \leq \parallel f^{k}(kT)\parallel$, $\forall
k\in\mathbb{N}$, the estimation :
\begin{equation}\label{eq:44}
\mid\parallel f \mid\parallel \leq \parallel f_{0}\parallel
\end{equation}
Now if \ $f, g \in C[0, +\infty; X_{r}]$ \ are two solutions of
(\ref{eq:43}), property (\ref{eq:22}) of \ $Q$ gives :
\begin{displaymath}
\parallel(f-g)(t)\parallel\leq 2Cr\int_{0}^{t}\parallel(f-g)(s)\parallel ds \ \ t\geq0
\end{displaymath}
which gives by Gronwall's Lemma \ $f=g$ \ and proves the uniqueness.\\
This completes the proof of Theorem \ref{t:4.3}. $\blacksquare$
\vskip 10pt\noindent \textbf{Acknowledgement} : The authors thank
A.D. Rendall for helpful comments and suggestions. This work was
supported in part by the Volkswagen\textbf{Stiftung}, Federal
Republic of Germany.

\end{document}